\documentclass[showpacs,twocolumn,pra]{revtex4}
\usepackage{amsmath}
\usepackage{amssymb}
\usepackage{graphicx}
\usepackage{subfigure}

\begin{document}

\title{Entanglement generation by collisions of quantum solitons in the
Born's approximation}
\author{Maciej Lewenstein$^{1,2}$ and Boris A. Malomed$^{3}$}
\affiliation{$^{1}$ICFO--Institut de Ci\`encies Fot\`oniques, E-08860 Castelldefels
(Barcelona), Spain\\
$^{2}$ICREA-- Instituci\'{o} Catalana de Recerca i Estudis Avan\c{c}ats,
E-08010 Barcelona, Spain\\
$^{3}$Department of Physical Electronics, School of Electrical Engineering,
Faculty of Engineering, Tel Aviv University, Tel Aviv 69978, Israel}

\begin{abstract}
We present analytic expressions describing generation of the entanglement in
collisions of initially uncorrelated quantum solitons. The results, obtained
by means of the Born's approximation (for fast solitons), are valid for both
integrable and non-integrable quasi-one-dimensional systems supporting
soliton states.
\end{abstract}

\maketitle

\section{Introduction}

Entanglement is a fundamental property of bipartite quantum systems \cite%
{horo}. Apart form being a major resource for quantum-information
techniques, entanglement exhibits itself perhaps in the most
spectacular form in the breakdown of Bell's inequalities and
Einstein-Podolsky-Rosen ``paradox". In experiments, strongly
entangled states, and in particular those exhibiting ``nonlocality",
are typically created with microscopic particles produced by the
same source, or interacting prior to the detection, such as pairs of
photons \cite{aspect,zeilinger} or ions \cite{blatt}. In their
original paper \cite{EPR}, Einstein, Podolsky and Rosen considered
two particles whose ordinary degrees of freedom, center-of-mass
positions and relative momenta, were correlated. This situation is
realized with the entanglement of Gaussian states of light
\cite{kimble} or atomic ensembles \cite{polzik}.

Particularly interesting is the possibility to generate entanglement
between \emph{macroscopic} (or mesoscopic) objects that may be
transmitted over long distances. In this connection,
\textit{solitons}, i.e., stable solitary waves that propagate
without distortion \cite{book}, may be considered as possible robust
quantum-information carriers. In particular, the solitons as
collective excitations in nonlinear media (unlike ions or other
material objects) may be created with a desirable shape and
effective mass, and admit a much greater degree of control by means
of various ``management" techniques \cite{management}.

The objective of the present work is to analyze the possibility of the
creation of entanglement between solitons by means of collisions between
them. To the best of our knowledge, this problem was not considered before
(except for a preliminary version of the present work \cite{boris}).
However, a very recent work \cite{Hsinchu} has presented a rigorous proof of
the generation of entanglement between constituent solitons in oscillating
two- and three-soliton bound states, in the integrable model based on the
NLS (nonlinear Schr\"{o}dinger) equation with the cubic nonlinearity. It is
relevant to note that multi-soliton solutions of the integrable NLS
equations are actually unstable against slow separation, hence the
entanglement, generated by the interactions between the solitons while they
remained bound, may be kept after the separation.

Natural candidates for the study of the collision-induced entanglement are
matter-wave solitons, that can be built of ultracold bosonic atoms, which
form the Bose-Einstein condensate (BEC) \cite{stringari}, and photonic
solitons in nonlinear optical waveguides \cite{book,Progress}. The studies
of BECs has led to the creation of dark \cite{maciek-sengstock,denschlag}
and bright \cite{khaykovich,Randy,cornish} solitons in trapped Bose-Einstein
condensates (with repulsive and attractive interactions, respectively).
Recently, the observation of stable bright-dark soliton pairs has been
reported too \cite{sengstock}. In addition, similarities to nonlinear optics
\cite{book} have triggered the interest in theoretical studies of discrete
(lattice) BEC solitons \cite{smerzi,anna}, and have also led to the seminal
observation of gap solitons, i.e., robust localized matter-wave packets
supported by the interplay of repulsive interactions and an effective
negative mass of collective excitations in the condensate, induced by the
periodic optical-lattice potential \cite{Markus}. The analysis reported
below suggests that the matter-wave solitons in BEC with attractive
inter-atomic interactions have the best potential for the generation of the
entanglement through collisions; in particular, the velocity of the moving
solitons, which is the crucial parameter, which determines the
collision-induced entanglement (see below), can be easily controlled for BEC
solitons.

While most of the previous studies of matter-wave solitons were concentrated
on their classical and mean-field aspects, more recently considerable
interest has been devoted to the role of thermal noise~\cite{muryshev,anna}
and quantum fluctuations. The latter may cause filling up of the
dark-soliton's core through the quantum depletion process, as was predicted
using the Bogoliubov-de Gennes (BdG) equations~\cite{dziarmaga}. Making use
of the discrete NLS equation and the time-evolving block-decimation
algorithm~\cite{vidal} in the framework of the Bose-Hubbard model, it was
confirmed that quantum effects lead to the filling in of dark BEC solitons,
and it has been demonstrated that collisions between them become inelastic~%
\cite{carr}. The BdG approach and its generalizations were also employed to
study the solitons' stability~\cite{yulin,Moti,Merhasin} and excitations
caused by opening~of the trap \cite{castin}. In the latter work, exact
Lieb-Liniger solutions were also used, to calculate internal correlation
functions of positions of the particles. A noisy version of standing bright
solitons was studied by means of the exact diagonalization and quantum
Monte-Carlo method~\cite{juha}.

Being typically associated with integrable 1D classical models, stable
solitons are known too in non-integrable systems, including multidimensional
ones. One of remarkable properties of solitons is their stability to
perturbations. The stability can be extended to quantum settings, in which
the mean-field description admits quantum solitons in the semi-classical
form. As mentioned above, the robustness of mean-field solitons against
quantum fluctuations (including finite-temperature effects) was studied,
using the time-dependent Hartree-Fock-Bogoliubov equations, for both
ordinary matter-wave solitons, supported by attractive interactions between
atoms \cite{Moti}, and for gap solitons \cite{Merhasin}. Although under
extreme conditions the quantum fluctuations may split a mean-field soliton
\cite{Moti}, it has been concluded that, in a broad range of parameters
relevant to the experiments, the matter-wave solitary waves predicted by the
mean-field description (i.e., found as stable localized solutions to the
Gross-Pitaevskii equation, GPE \cite{GPE}) are completely robust objects --
in fact, in perfect agreement with the experimental observations of these
solitons \cite{Randy}-\cite{Markus}. As concerns the relation between the
mean-field and quantum descriptions of localized objects in BEC, it is
relevant to mention that an alternative derivation of the GPE from a
consistent many-body quantum theory, based on the variational approach in
the multi-configurational space, was recently presented in Refs. \cite{Alon}.

In this work, we analyze the collision-induced generation of the
entanglement in pairs of fast solitons for a class of equations of the NLS
type by means of the Born approximation, which is valid when the kinetic
energy of the moving objects (solitons) is much larger than the potential of
the interaction between them, hence the interaction may be treated as a
perturbation \cite{LL} (therefore, this is the case opposite to that studied
in Ref. \cite{Hsinchu}, where the entanglement was considered between bound
solitons created with zero relative velocity). We aim to present simple
analytical expressions for the collision-generated entanglement of two
quantum solitons, which are valid for generic quasi-1D systems -- integrable
or not -- that admit soliton solutions. Another approach is possible for
nearly quiescent solitons, when the entanglement-generating perturbation is
the weak interaction between them, assuming that the distance between the
solitons is large enough. Recently, this approach was developed in Ref. \cite%
{Reznik} for kinks in the sine-Gordon equation. In that model, the kinks
were maintained in the quiescent state by boundary conditions, as the model
was defined in a domain of a finite size.

The rest of the paper is organized as follows. In Section II, we introduce
the model, starting with the known system of classical equations of motion
for a pair of well-separated solitons \cite{Progress}, and then proceed to
the respective quantum system. We confine the analysis to the basic case of
the pair of symmetric solitons with equal amplitudes. Mismatch between the
amplitudes makes the interaction effectively incoherent \cite{Gordon}, thus
suppressing the entanglement generation (we estimate the size of the
mismatch up to which it may be neglected). The initial quantum state,
corresponding to far separated solitons, is taken as a product of two
independent wave packets. In Section III, the calculation of the correction
to this factorized state, generated by the collision between the fast
solitons, is performed in an analytical form, by means of the
above-mentioned Born approximation. The collision-induced correction to the
wave function features \emph{explicit entanglement} in terms of two relevant
degrees of freedom, \textit{viz}., the distance between the solitons and
their relative phase, $r$ and $\chi $. The results are reported for the
initial condition of two types: a more sophisticated one, with the Gaussian
localization of $\chi $ around a definite value, and also for a simple
phase-independent initial distribution. Both types of the initial quantum
states may be realized in the experiment, under different specific
conditions. The paper is concluded by Section IV, where, in particular, we
discuss the robustness of the predicted entanglement against external noise,
and possible extensions of the work.

\section{The model}

\subsection{The classical soliton pair}

For the condensate with attractive interactions between atoms, the scaled
form of the GPE in the free 1D space, which describes the BEC in the
mean-field approximation, is tantamount to the integrable NLS equation \cite%
{GPE}, i.e.,%
\begin{equation}
iu_{t}+(1/2)u_{xx}+\left\vert u\right\vert ^{2}u=0.  \label{NLS}
\end{equation}%
The commonly known soliton solution to Eq. (\ref{NLS}) is
\begin{equation}
u_{\mathrm{sol}}=\eta \,\mathrm{sech}\left[ \eta (\left( x-\xi (t)\right) %
\right] \,\exp \left[ i\phi (t)+icx\right] \,,  \label{eta}
\end{equation}%
where $\eta $ is the soliton's amplitude, $d\xi /dt=c$ its velocity, and
\begin{equation}
-d\phi /dt=\left( c^{2}-\eta ^{2}\right) /2  \label{freq}
\end{equation}
the intrinsic frequency.

More realistic forms of the GPE in 1D include various terms which break the
integrability of the NLS equation. In particular, the full equation must
include the axial trapping potential, but, as concerns interactions between
solitons, the axial potential is not a crucially important factor, according
to the available experimental results \cite{Randy} and theoretical analysis
\cite{Brand,Lev}. Another physically relevant feature that may affect
soliton-soliton collisions is a quintic nonlinear term. It may account for
three-body collisions in the condensate, provided that they are lossless
\cite{Tomio}, but a more general (in fact, universal) source of the quintic
term is the deviation of the condensate loaded into a cigar-shaped trap from
the one-dimensionality. This term can be derived by means of a perturbative
analysis \cite{non1D}, or by the expansion of a more general equation, that
takes into regard the underlying three-dimensionality via the nonpolynomial
nonlinearity \cite{Luca}. The universal quintic term always corresponds to
\emph{self-attraction} (irrespective of the sign of the binary interactions
between atoms), with the coefficient in front of it proportional to the
transverse-confinement frequency and square of the collisional scattering
length. By means of straightforward rescalings, the equation with the
combination of attractive cubic and quintic terms can be cast in the form
that contains no free parameters,%
\begin{equation}
iu_{t}+(1/2)u_{xx}+\left\vert u\right\vert ^{2}u+|u|^{4}u=0.  \label{CQ}
\end{equation}%
Although the cubic-quintic (CQ) NLS equation (\ref{CQ}) is not integrable,
its exact soliton solutions are well known \cite{Bulgaria}, the entire
family being stable \cite{Seva}:%
\begin{equation}
\tilde{u}_{\mathrm{sol}}=\frac{\sqrt{2}\eta \exp \left[ i\phi (t)+icx\right]
}{\sqrt{\sqrt{1+\left( 8/3\right) \eta ^{2}\cosh \left[ 2\eta \left( x-\xi
(t)\right) \right] }-1}},  \label{tilde}
\end{equation}%
where $\phi (t)$ and $\xi (t)$ have the same meaning as in Eq. (\ref{eta}).
The CQ model based on equation (\ref{CQ}) finds other physical realizations
in nonlinear optics \cite{optics}.

The analysis which treats the interaction between well separated
quasi-classical (mean-field) solitons (\ref{eta}) with common amplitude $%
\eta $ as a perturbation leads to the the conclusion that they may be
considered as a pair of quasi-particles with an effective Hamiltonian \cite%
{Progress},%
\begin{equation}
H_{\mathrm{tot}}=(1/2)\left( \dot{\rho}^{2}-\dot{\theta}^{2}+\dot{r}^{2}-%
\dot{\chi}^{2}\right) -e^{-\left\vert r\right\vert }\cos \chi ,
\label{total}
\end{equation}%
where the overdot stands for the differentiation with respect to rescaled
time $\tau \equiv 2\sqrt{2}\eta ^{2}t$, the normalized distance between the
solitons and center-of-mass coordinate are defined as
\begin{equation}
r\equiv \eta (\xi _{1}-\xi _{2}),~\rho \equiv \eta \left( \xi _{1}+\xi
_{2}\right) ,  \label{r}
\end{equation}%
while the relative and overall phases of the soliton pair are%
\begin{equation}
\chi \equiv \phi _{1}-\phi _{2},~\theta \equiv \phi _{1}+\phi _{2}-\eta
^{2}t.  \label{chi}
\end{equation}%
A noteworthy fact, which is obvious in expression (\ref{total}), is that the
effective mass corresponding to the phase degrees of freedom in the
two-soliton set is \emph{negative} \cite{Seva2}.

Actually, Hamiltonian (\ref{total}) is universal, applying to any model
which supports exponentially localized solitons with an intrinsic phase
degree of freedom \cite{me}. In particular, this Hamiltonian governs the
interaction between solitons (\ref{tilde}) of the NLS equation with the CQ
nonlinearity, Eq. (\ref{CQ}), up to a proper rescaling of coefficients \cite%
{Lev}.

\subsection{Quantization}

Proceeding to the quantum version of the model considered above, we treat,
as usual \cite{Rajaraman}, each soliton as a quantum particle with two
degrees of freedom, the position and phase (in the experimentally relevant
situation, effects of quantum fluctuations around the quasi-classical shape
of the solitons may be negligible for matter-wave solitons, as discussed
above). Thus, the quantum counterpart of classical Hamiltonian \ref{total}
gives rise to the following linear Schr\"{o}dinger equation for the wave
function of the soliton pair, $\Psi ^{\left( \mathrm{tot}\right) }$, which
depends on the total set of four degrees of freedom describing the pair:%
\begin{eqnarray}
i\Psi _{T}^{\left( \mathrm{tot}\right) } &=&-(1/2)\left[ \Psi _{rr}^{\left(
\mathrm{tot}\right) }-\Psi _{\chi \chi }^{\left( \mathrm{tot}\right) }+\Psi
_{\rho \rho }^{\left( \mathrm{tot}\right) }-\Psi _{\theta \theta }^{\left(
\mathrm{tot}\right) }\right]  \notag \\
&&-\varepsilon ~e^{-\left\vert r\right\vert }\left( \cos \chi \right) \Psi
^{\left( \mathrm{tot}\right) },  \label{Psi-tot}
\end{eqnarray}%
where $T\equiv \hbar \tau $ and $\varepsilon \equiv 1/\hbar ^{2}$, with $%
\hbar $ the renormalized Planck's constant [measured in scaled units in
which Hamiltonian (\ref{total}) was written]. The reduced form of Eq. (\ref%
{Psi-tot}) for the wave function which depends only on the relative
variables, $r$ and $\chi $, is%
\begin{equation}
i\Psi _{T}=-\left[ \frac{1}{2}\left( \Psi _{rr}-\Psi _{\chi \chi }\right)
+\varepsilon ~e^{-\left\vert r\right\vert }\cos \chi \right] \Psi .
\label{Psi}
\end{equation}

Our objective is to analyze collisions between rapidly moving quantum
solitons, with large relative momentum $K_{0}$. The initial state is taken
as a naturally expected non-entangled factorized one, centered around
definite initial values of the dynamical variables, $\pm \xi _{0}$ and $\pm
\chi _{0}$:%
\begin{gather}
\Psi _{0}^{(\mathrm{tot})}\left( \xi _{1},\xi _{2},\phi _{1},\phi
_{2}\right) =  \notag \\
\equiv \exp \left[ -\frac{\eta _{0}^{2}\left( \xi _{1}-\xi _{0}\right) ^{2}}{%
2\Xi ^{2}}-\frac{\left( \phi _{1}-\chi _{0}\right) ^{2}}{2\Phi ^{2}}\right]
e^{iK_{0}\eta \left( \xi _{1}-\xi _{0}\right) }  \notag \\
\times \exp \left[ -\frac{\eta _{0}^{2}\left( \xi _{2}+\xi _{0}\right) ^{2}}{%
2\Xi ^{2}}-\frac{\left( \phi _{2}+\chi _{0}\right) ^{2}}{2\Phi ^{2}}\right]
e^{-iK_{0}\eta \left( \xi _{2}+\xi _{0}\right) }~.  \label{product}
\end{gather}%
This state assumes equal widths of the wave packets for the two solitons, $%
\Xi $ and $\Phi $, with the pair's center of mass set at $x=0$, and initial
separation $2\xi _{0}\equiv \xi _{1}(T=0)-\xi _{2}(T=0)$. The mean value of
the initial overall phase is also fixed to be zero, while the initial phase
difference between the solitons is $2\chi _{0}\equiv \phi _{1}(T=0)-\phi
_{2}(T=0)$. Finally, initial state (\ref{product}) can be written in terms
of the variables defined in Eqs. (\ref{r}) and (\ref{chi}),
\begin{gather}
\Psi _{0}^{(\mathrm{tot})}\left( \xi _{1},\xi _{2},\phi _{1},\phi
_{2}\right) =e^{iK_{0}\left( r-r_{0}\right) }  \notag \\
\times \exp \left[ -\frac{\left( r-r_{0}\right) ^{2}+\rho ^{2}}{4\Xi ^{2}}-%
\frac{\left( \chi -\chi _{0}\right) ^{2}+\theta ^{2}}{4\Phi ^{2}}\right] ,
\label{product2}
\end{gather}%
where and $r_{0}\equiv 2\eta \xi _{0}$.

To conclude the formulation of the model, it is relevant to consider
the possibility of the excitation of an intrinsic mode (IM) in the
colliding solitons (if the IM exists). A known principle is that
solitons of integrable equations do not support IMs, but a
nonintegrable model may feature an IM. In particular, exactly one IM
is exists in excited states of solitons (\ref{tilde}) of the CQ NLS
equation, with the self-focusing sign of both nonlinear terms
\cite{Seva}. A possibility of using solitons' IMs as carriers of
quantum information was proposed in Ref. \cite{Reznik}. However, in
the case of the collision between fast solitons, which is considered
below, the excitation of the IM, as well as generation of nonsoliton
modes (``radiation"), may be neglected, in the lowest approximation,
simply because the intensity of these effects is inversely
proportional to the square of the collision velocity
\cite{Progress}.

\section{Generation of entanglement by collisions between solitons}

\subsection{The Born's approximation}

Scattering solutions to reduced equation (\ref{Psi}) are generated by the
incident wave, $\Psi (r\rightarrow \infty )=\exp \left[ iKr+i\kappa \chi
-i\left( K^{2}-\kappa ^{2}\right) T/2\right] $, where $\kappa $ is an
integer. The full solution is sought for as
\begin{equation}
\Psi \left( T,r,\chi \right) =V\left( r,\chi \right) \exp \left[ -i\left(
K^{2}-\kappa ^{2}\right) T/2\right] ,  \label{T}
\end{equation}%
with the stationary part of the wave functions obeying the following
equation,%
\begin{equation}
V_{rr}-V_{\chi \chi }+\left[ 2\varepsilon ~e^{-\left\vert r\right\vert }\cos
\chi +\left( K^{2}-\kappa ^{2}\right) \right] V=0.  \label{Phi}
\end{equation}%
As said above, our basic assumption is that we consider the collision
between fast solitons, i.e., $K^{2}$ is a large parameter in comparison with
$\varepsilon $ ($\kappa ^{2}$ may be large too). In other words, the kinetic
energy of the relative motion is much larger than the potential of the
soliton-soliton interaction. In fact, the two-particle Hamiltonian (\ref%
{total}) can be used for the description of collisions between solitons only
in this case; otherwise, one cannot neglect deformation of the solitons in
the course of the collision, as well as the excitation of the IM, if it
exists in the solitons, and the generation of the radiation modes.

We apply the Born's approximation \cite{LL}, looking for a solution to Eq. (%
\ref{Phi}) as
\begin{equation}
V(r,\chi )=\left[ 1+\varphi \left( r,\chi \right) \right] \exp \left(
iKr+i\kappa \chi \right) ,  \label{Born}
\end{equation}%
where perturbation $\varphi $, which is assumed to be a slowly varying
function of $r$ in comparison with $\exp \left( iKr\right) $, obeys the
simplified equation: $-iK\varphi _{r}+(1/2)\varphi _{\chi \chi }+i\kappa
\varphi _{\chi }=\varepsilon ~e^{-\left\vert r\right\vert }\cos \chi $.
Solutions to this equation satisfying the necessary boundary condition, $%
\varphi (r=-\infty )=0$, can be readily found in an analytical form. The
outcome of the collision is determined by the asymptotic form of the
solution, which is found to be
\begin{gather}
\varphi (r\rightarrow +\infty ,\chi )=\left( i\varepsilon /K\right)  \notag
\\
\times \left[ e^{i\chi }e^{i\left( \kappa -1/2\right) r/K}+e^{-i\chi
}e^{-i\left( \kappa +1/2\right) r/K}\right] .  \label{solution}
\end{gather}

\subsection{Analysis of the collision-induced entanglement}

In the case of the fast collision, the overall variables, $\rho $ and $%
\theta $, may be treated as ``frozen" ones in the wave packet
generated by initial state (\ref{product2}). Then, it is natural to
decompose the initial state over the set of plane waves with respect
to
variables $r$ and $\chi $:%
\begin{gather}
\Psi _{0}^{(\mathrm{tot})}\left( \xi _{1},\xi _{2},\phi _{1},\phi
_{2}\right) =\exp \left( -\frac{\rho ^{2}}{4\Xi ^{2}}-\frac{\theta ^{2}}{%
4\Phi ^{2}}\right)   \notag \\
\times \left( \Xi \Phi /\pi \right) \int_{-\infty }^{+\infty
}dK\int_{-\infty }^{+\infty }d\kappa e^{iK\left( r-r_{0}\right) +\kappa
\left( \chi -\chi _{0}\right) }  \notag \\
\times \exp \left[ -\Xi ^{2}\left( K-K_{0}\right) ^{2}-\Phi ^{2}\kappa ^{2}%
\right] ~.  \label{decomp}
\end{gather}%
Note that the Gaussian distribution of angular wavenumber $\kappa $, which
must be integer, is valid for $\left\vert \kappa \right\vert \gg 1$, i.e., $%
\Phi \ll 2\pi $ [recall that $\Phi $ is the width of the initial
distribution of the phase variables introduced in Eqs. (\ref{product}) and (%
\ref{product2})].

Next, recombining wave packet (\ref{decomp}) with the collision-induced
perturbation of the wave function, as per Eqs. (\ref{Born}) and (\ref%
{solution}), and again making use of the fact that $K_{0}$ is large, we
arrive at an expression for the net change of the wave function which is
generated by the fast collision between the two solitons, in the first order
of the perturbation theory:
\begin{gather}
\delta \Psi \left( \xi _{1},\xi _{2},\phi _{1},\phi _{2}\right) \approx \exp
\left( -\frac{\rho ^{2}}{4\Xi ^{2}}-\frac{\theta ^{2}}{4\Phi ^{2}}\right)
\notag \\
\times \frac{i\sqrt{2}\varepsilon \Phi }{K_{0}\left( 4\Phi ^{4}+T^{2}\right)
^{1/4}}\exp \left[ i\left( K_{0}\left( r-r_{0}\right) -\frac{1}{2}%
K_{0}^{2}T\right) \right]   \notag \\
\times \sum_{+,-}e^{\pm i\chi }\exp \left[ -\frac{\Phi ^{2}\left( \chi -\chi
_{0}\pm r/K_{0}\right) ^{2}}{4\Phi ^{4}+T^{2}}+i\Omega _{\pm }\right] ,
\label{final}
\end{gather}%
with phases shifts%
\begin{equation}
\Omega _{\pm }\equiv \frac{1}{2}\left( \tan ^{-1}\left( \frac{T}{2\Phi ^{2}}%
\right) -\frac{T\left( \chi -\chi _{0}\pm r/K_{0}\right) ^{2}}{4\Phi
^{4}+T^{2}}\right) .
\end{equation}%
A nontrivial feature of expression (\ref{final}), which represents the \emph{%
entanglement proper}, is the combination of two harmonics, $\exp \left( \pm
i\chi \right) $, multiplied by the Gaussian factors, whose maxima are
located along two spirals in the plane of $\left( r,\chi \right) $: $r_{\max
}=\mp K_{0}\left( \chi -\chi _{0}\right) $. These maxima also determine the
correlation between variables $r$ and $\chi $, which are an inherent part of
the entanglement. The width of the maxima gradually spreads out with the
growth of time ($T$), proportionally to $\sqrt{4\Phi ^{4}+T^{2}}$, as does
any coherent state evolving in the free space.

It is relevant here to estimate a dephasing effect of a possible mismatch
between amplitudes of the colliding solitons, $\Delta \eta $. It may be
estimated  through the respective change of the relative phase, $\Delta \chi
\sim \eta \Delta \eta \Delta t_{\mathrm{coll}}\sim \Delta \eta /c$,
generated by the mismatch during the collision time, $\Delta t_{\mathrm{coll}%
}\sim 1/\left( \eta c\right) $ [see Eqs. (\ref{NLS})-(\ref{freq})]. The
dephasing is negligible if its size is small in comparison with the
perturbation amplitude in expression (\ref{final}), i.e.,
\begin{equation}
\Delta \eta /c\ll \varepsilon /K_{0}.  \label{mismatch}
\end{equation}%
Currently available sophisticated experimental methods for the creation of
solitons in BEC \cite{Randy2}, as well as various theoretically elaborated
schemes of matter-wave soliton lasers \cite{laser}, make it possible to
generate nearly identical solitons, with a sufficiently small difference
between their amplitudes.

\subsection{The initial phase-uniform state}

Instead of initial wave packet (\ref{product}), we can take one uniformly
spread over phases $\phi _{1}$ and $\phi _{2}$, which corresponds to an
experimental situation in which the initial phases of solitons are not
controlled. Then, modifying expression Eq. (\ref{product2}) accordingly,
decomposition (\ref{decomp}) is replaced by its simplified counterpart:
\begin{gather}
\Psi _{0}^{(\mathrm{tot})}\left( \xi _{1},\xi _{2}\right) =\exp \left( -%
\frac{\rho ^{2}}{4\Xi ^{2}}\right)   \notag \\
\times \left( \Xi /\sqrt{\pi }\right) \int_{-\infty }^{+\infty
}dKe^{iK\left( r-r_{0}\right) }\exp \left[ -\Xi ^{2}\left( K-K_{0}\right)
^{2}\right] .  \notag
\end{gather}%
Recombining this with the result of the Born's approximation, as per Eq. (%
\ref{solution}), gives rise to the following expression for the
collision-induced perturbation of the wave function:
\begin{gather}
\delta \Psi \left( \xi _{1},\xi _{2},\chi \right) =\exp \left( -\frac{\rho
^{2}}{4\Xi ^{2}}\right) \cos \chi   \notag \\
\times \frac{2i\varepsilon }{K_{0}}\exp \left[ i\left( K_{0}\left(
r-r_{0}\right) -\frac{1}{2}K_{0}^{2}T\right) \right]   \label{no-phase-final}
\end{gather}%
Although this result is much simpler than the one given by Eq. (\ref{final}%
), in the case when the solitons' phases were initially allocated certain
values, it is nontrivial too, demonstrating the dependence on relative phase
$\chi $ of the perturbation generated, in the course of the collision, by
the phase-independent initial wave function. Actually, this feature may be
regarded as the simplest manifestation of the collision-induced
entanglement. The correlation properties of the entanglement in this
approximation are obvious, being, as a matter of fact, determined by factor $%
\cos \chi $.

\section{Conclusion}

In this work, our aim was to present a proof of the principle that two
initially uncorrelated quantum (actually, semi-classical) solitons may get
entangled through the collision between them. To this end, we have
restricted the analysis to macroscopic coordinates of the solitons, \textit{%
viz}., the position and phase, which may be justified for
matter-wave solitons in BEC. In this sense, they are regarded as
Einstein-Podolsky-Rosen particles with the additional internal
degree of freedom (the phase). We have presented simple analytic
expressions characterizing the resulting entangled states,
represented by collision-induced corrections to initial factorized
wave functions of the soliton pair. The results were obtained by
means of the Born's approximation, which is valid for the collision
between fast soliton. The entanglement and quantum correlations are
described in terms of the relative position and phase of the
solitons, and the (conserved) total momentum and ``angular momentum"
(the latter is the canonical momentum conjugate to the total phase).
A simple but noteworthy effect is that the collision of solitons
with the completely uncertain (delocalized) relative phase leads to
a partial phase localization. The predicted results are expected to
be directly relevant to matter-wave solitons in BEC formed by atoms
with attractive interactions and loaded into a nearly 1D trapping
potential. A realization in terms of optical solitons may be
possible too, in principle.

The entanglement predicted by the above analysis is weak, as the
consideration was limited to the Born approximation, i.e., the first
approximation of the perturbation theory. However, under more general
conditions, the entanglement may reach significant levels (cf. Ref. \cite%
{Hsinchu} where the entanglement was analyzed by means of an exact quantum
solution for NLS solitons interacting at zero velocity). What is then a
possible advantage of using matter-wave solitons over individual particles,
likes photons or atoms? The obvious answer is that the solitons constitute
macroscopic, or at least mesoscopic objects, and thus exhibit completely
different properties in measurements. To some extent, they are similar to
atomic ensembles \cite{polzik}, but, in contrast to the latter, the solitons
represent coherent moving objects. In principle, the Bell's inequalities can
be tested with entangled solitons, and even though such test would not be
loophole-free, they would evidently exhibit different scaling and level of
errors, in comparison with previously studied systems.

As said above, one may attain stronger correlations by going beyond the
Born-approximation limit. However, in that regime the macroscopic
(mean-field) description may cease to be quantitatively correct -- in
particular, because the deformation of the slowly colliding solitons makes
the it irrelevant to use the description solely in terms of their collective
degrees of freedom. Then, it may be appropriate to consider quantum solitons
using the time-dependent GPE [or similar equation(s)] to describe the
coherent soliton states proper, and the BdG (Bogoliubov-de Gennes) equations
for quantum fluctuations around them \cite{GPE}, cf. the analysis of the
stability of matter-wave solitons against quantum fluctuations performed in
Refs. \cite{Moti} and \cite{Merhasin}. In particular, care should be taken
of the collisional excitation of an IM (intrinsic mode), if the solitons
support it. However, such a consideration is beyond the scope of this work.

Similar comments concern another important aspect of the studies of the
entanglement, \textit{viz}., the decoherence. In the present case, a part of
the decoherence is due to the excitation of many-body degrees of freedom
(such as BdG modes), rather than just quantum fluctuations of the solitons'
positions and phases. Nevertheless, in the case of the collision between
fast solitons we may restrict the attention to the collective degrees of
freedom of the solitons, and look at the respective covariance matrix,
similarly to the case of Gaussian states. Then, applying the entanglement
criteria for Gaussian states \cite{zoller,simon,giedke} makes it possible to
estimate the level of the entanglement of the present (ideal) state, as well
as the maximum allowance of the contribution to the decoherence from the
neglected many-body modes.

The estimate of the robustness of the entanglement is particularly simple
for the solution corresponding to Eq. (\ref{no-phase-final}), because the
degrees of freedom corresponding to the distance and phase shift between the
solitons separate, hence the wave function of the two-soliton state can be
written, in terms of dynamical variables (\ref{r}) and (\ref{chi}), as [cf.
Eqs. (\ref{final}), (\ref{no-phase-final})]

\begin{gather}
|\Psi (\rho ,r,\chi )\rangle =|\Psi (\rho )\rangle |\Psi (r)\rangle   \notag
\\
\times \left[ |0\rangle +\zeta \left( |1\rangle +|-1\rangle \right) \right] ,
\label{entang}
\end{gather}%
where states of the relative phase ($\chi $), $|0\rangle ,|\pm
1\rangle $, correspond to harmonics with the respective numbers, and
$\zeta \equiv \varepsilon /\left( 2\sqrt{2}K_{0}\right) $. The phase
part of state (\ref{entang}) is obviously entangled, since its
partial transpose has three
negative eigenvalues, namely, $-\zeta ,-\zeta ,-\zeta ^{2}$ \cite{horodecki}%
. Therefore, the state remains entangled after adding arbitrary
decoherence (noise) whose density matrix has an operator norm
smaller than $\zeta $. This actually means that one can add ``white
noise" with norm $\zeta ^{\prime }<\zeta $ to state (\ref{entang}),
and retain its non-positive partial transpose property. As concerns
the dephasing due to a possible difference in the amplitudes of the
colliding solitons, the
estimate performed above demonstrates that it is negligible under condition (%
\ref{mismatch}).

A straightforward extension of the present analysis, still within the
framework of the description based on the mean-field collective degrees of
freedom and the Born's approximation, may be elaborated for the description
of the generation of entanglement by collisions between two-dimensional
solitons supported by a quasi-1D potential in the self-attractive BEC, as
proposed (in the mean-field approximation) in Refs. \cite{BBB}. Another
extension, also possible in the framework of the quasi-particle approach, is
to consider entanglement induced by collisions between gap solitons in a
self-repulsive BEC loaded into an optical-lattice potential. It is known
that such solitons may be mobile in 1D and 2D geometries, with a \emph{%
negative} effective mass \cite{HS}. The interaction potential in the
corresponding Hamiltonian is expected to be different from the above
expression (\ref{total}), namely, being $\sim \exp \left( -|r|\right) \cos
(qr)\cos \chi $, with some wavenumber $q$, due to the oscillatory shape of
tails of the gap solitons.

B.A.M. appreciates hospitality of ICFO (Barcelona) and a useful discussion
with B. Reznik and S. Marcovitch. We acknowledge Spanish MEC/MINCIN projects
TOQATA (FIS2008-00784) and QOIT (Consolider Ingenio 2010), ESF/MEC project
FERMIX (FIS2007-29996-E), EU STREP project NAMEQUAM, ERC Advanced Grant
QUAGATUA, Alexander von Humboldt Foundation Senior Research Prize, and grant
No. 149/2006 from the German-Israel Foundation.

\end{document}